\documentclass[submission
]{eptcs}
\providecommand{\event}{TLLA 2022} 
\usepackage{breakurl}             
\usepackage{underscore}           

\usepackage[T1]{fontenc} 
\usepackage[utf8]{inputenc} 
\usepackage{lmodern} 
\usepackage[english]{babel} 
\usepackage{amssymb}  
\usepackage{stmaryrd} 
\usepackage{amsthm}
\usepackage{amsmath}
\usepackage{comment}
\usepackage{ebproof}
\usepackage{tkz-graph}
\usepackage{tikz}
\usepackage{tikz-cd}
\usetikzlibrary {positioning}
\usepackage{cmll}
\usepackage{mathrsfs}
\usepackage{marvosym}
\usepackage{cancel}
\usepackage{pst-solides3d}
\usepackage{MnSymbol}
\usetikzlibrary{matrix,arrows,decorations.pathmorphing,calc}
\usepackage[top=1in, bottom=1.25in, left=1in, right=1in]{geometry}

\newtheorem{corollary}{Corollary}

\newtheorem{proposition}{Proposition}
\newtheorem{remark}{Remark}

\newtheorem{definition}{Definition}

\newcommand{\N}{\mathbb{N}}

\usepackage{setspace}

\newcommand\seq\vec 
\newcommand{\st}{\mid} 
\newcommand{\set}[1]{\{#1\}}

\newcommand\bpf{\begin{proof}}
\newcommand\epf{\end{proof}}

\newcommand{\Pfinstar}[1]{\mathscr{P}_\mathrm{fin}^*(#1)}
\newcommand{\web}[1]{\lvert #1\lvert}
\newcommand{\homology}[2]{\mathcal{H}_{#1}(#2)}
\newcommand{\cs}[1]{\mathrm{cs}(#1)}
\newcommand{\simplices}[1]{\mathcal{S}#1}

\newcommand{\barI}[1]{\mathscr{I}\!#1}
\newcommand{\barsubd}[1]{\mathbf{bs}(#1)}

\newcommand{\cutelim}{\rightsquigarrow}

\title{Denotational semantics driven simplicial homology?}
\author{Davide Barbarossa\thanks{Thanks to Thomas Ehrhard who contributed to this work. Thanks also to Giulio Manzonetto and Lorenzo Tortora de Falco for instructive discussions.}
\institute{Dipartimento di Informatica: Scienza ed Ingegneria\\
Università di Bologna
\\
Bologna, Italia}
\email{davide.barbarossa@unibo.it}
}
\def\titlerunning{Den.\ sem.\ and homology}
\def\authorrunning{Davide Barbarossa}
\begin{document}
\maketitle

\paragraph{About this article}
This article was presented at TLLA 2022 and the present version is the same but with a slightly longer and clearer abstract.

\begin{abstract}
One of the main preoccupations of mathematical logic is the mathematical study of mathematical proofs.
Typically, one studies their internal structure by means of the cut-elimination procedure.
One also thinks to be giving a ``meaning'' to the proofs of a fixed proof-system admitting a cut-elimination procedure by giving a denotational semantics for it,
that is, by finding invariants to the dynamics.
For a fixed proof-system -- say some fragment of Linear Logic -- we propose, instead, to study its proofs by looking at their external structure: in fact, Linear Logic's coherent semantics interprets the proofs of a given formula $A$ as faces of an abstract simplicial complex, thus allowing us to see the set of the (interpretations of the) proofs of $A$ as a \emph{geometrical space}, not just a \emph{set}.
This point of view has never been really investigated.
A space coming from the coherent semantics is, by construction, of a particular kind, namely it is a clique complex, and thus its faces do not in general correspond to proofs of $A$.
But actually, for any ``webbed'' denotational semantics -- say the relational one --, it suffices to down-close the set of (the interpretations of the) proofs of $A$ in order to give rise to an abstract simplicial complex whose faces do correspond to proofs of $A$.
Since this space comes triangulated by construction, a natural geometrical property to consider is its homology.
However, we immediately stumble on a problem: if we want the homology to be invariant w.r.t.\ to some notion of type-isomorphism, we are naturally led
to consider the homology functor acting, at the level of morphisms, on ``simplicial relations'' rather than simplicial maps as one does in topology.
The task of defining the homology functor on this modified category can be achieved by considering a very simple monad, which is almost the same as the power-set monad;
but, doing so, we end up considering not anymore the homology of the original space, but rather of its transformation under the action
of the monad.
Does this transformation keep the homology invariant ? Is this transformation meaningful from a geometrical or logical/computational point of view ?
We do not know the answers to these questions yet, so this work only sets up a curious question;
nevertheless we think that the perspective considering the whole of the proofs of a given formula as a  geometrical space -- not only a set -- could be of interest.
\end{abstract}

\section{A natural question}

 An \emph{abstract simplicial complex} $X$ (asc for short) is the data of a (at most countable) set $\web X$ and a $\subseteq$-initial segment $\mathcal S X$ of $\Pfinstar{\web{X}}$ s.t.\ $\mathcal S X \supseteq \set{\set{a} \mid a\in \web X}$.
 The set $\web X$ is called the \emph{web} of $X$ and its elements are the \emph{vertices} of $X$.
 The elements of $\simplices X$ are the \emph{simplices} of $X$.
 The subsets $y$ of a simplex $x$ of $X$ are the \emph{faces} of $x$ (thus, faces are simplices).
 The elements of $\Pfinstar {\web X} - \simplices X$ are called the \emph{non-simplices} of $X$.
Famous asc's are $\Delta^n:=(\set{0,\dots,n},\mathscr{P}^*(\set{0,\dots,n}))$ and $S^{n-1}:=(\set{0,\dots,n},\simplices{\Delta^n}-\set{0,\dots,n})$.
An asc $X$ can be \emph{geometrically realized} as a topological space 
$\mathrm{Geo}(X)$ inside $\mathbb{R}^{\mathrm{Card}(\web X)}$ via a standard and well known construction \cite{Hatcher}.
For example, $\Delta^n$ becomes the \emph{standard $n$-simplex} ($\Delta^1$ the segment between $(1,0)$, $(0,1)$ in $\mathbb{R}^2$, $\Delta^2$ the filled triangle of the vertices $(1,0,0)$, $(0,1,0)$, $(0,0,1)$ in $\mathbb{R}^3$ etc.), $S^{n-1}$ becomes the border of $\mathrm{Geo}(\Delta^n)$, homeomorphic to the $(n-1)$-sphere.
The natural notion of morphism between asc's is that of simplicial maps:
maps $f:\web X \, \to \web Y$ sending simplices of $X$ to simplices of $Y$.
We denote this category with ASC.

\begin{figure}[t]
    \centering
    \begin{minipage}{0.2\textwidth}
        \centering
         \begin{footnotesize}\begin{tikzcd}
            T_1 \arrow[r,-]\arrow[d,-] 
            & T_2 \arrow[d,-] \\
            F_2 \arrow[r,-] & F_1
         \end{tikzcd}\end{footnotesize}
        \caption{$\llbracket\mathrm{Bool}^{\& 2}\rrbracket$}\label{fig:Bool2}
    \end{minipage}
    \begin{minipage}{0.35\textwidth}
        \centering
        \begin{scriptsize}\begin {tikzpicture}
          \node (t1) at (0,0) {$T_1$};
          \node (t3) at (0.4,1.7) {$T_3$};
          \node (t2) [right = of t1] {$T_2$};
          \node (f2) at (-0.9,-0.7) {$F_2$};
          \node (f1) [right = of f2] {$F_1$};
          \node (f3) at (0.4,-2.4)  {$F_3$};
          \path (t3) edge (t1);
          \path (t3) edge (t2);
          \path (t3) edge (f2);
          \path (t3) edge (f1);
          \path (f3) edge (t1);
          \path (f3) edge (t2);
          \path (f3) edge (f2);
          \path (f3) edge (f1);
          \path (f2) edge (f1);
          \path (t2) edge (t1);
          \path (t1) edge (f2);
          \path (t2) edge (f1);
        \end{tikzpicture}\end{scriptsize}
        \caption{$\llbracket\mathrm{Bool}^{\& 3}\rrbracket$}\label{fig:Bool3}
    \end{minipage}
    \begin{minipage}{0.45\textwidth}
        \centering
          \begin{footnotesize}$
          \dfrac{
          \dfrac{
          \dfrac{
          \dfrac{}{\vdash 1}\mathrm{1}
          }{\vdash\mathrm{Bool}}\oplus_i
          \qquad
          \dfrac{
          \dfrac{}{\vdash 1}\mathrm{1}
          }{\vdash\mathrm{Bool}}\oplus_j
          }{\vdash\mathrm{Bool}^{\& 2}}\&
          \qquad
          \dfrac{
          \dfrac{}{\vdash 1}\mathrm{1}
          }{\vdash\mathrm{Bool}}\oplus_m
          }{\vdash\mathrm{Bool}^{\& 3}}\&$\end{footnotesize}
        \caption{The proofs $\pi_{ijm}$}\label{fig:piijm}
    \end{minipage}
\end{figure}

\emph{Coherent spaces} (\emph{cs}'s for short) \cite{girard1989proofs} are exactly the asc's s.t.\ any $\subseteq$-minimal non-simplex has cardinality $2$.
They are well known in geometry under the name of \emph{flag asc's}.  
We can interpret LL within the coherent semantics $\llbracket\cdot\rrbracket$, so a formula $A$ becomes a space $\mathrm{Geo}(\llbracket A \rrbracket)\subseteq\mathbb{R}^{\mathrm{Card}(\web{\llbracket A \rrbracket})}$ and its proofs, which are cliques of the coherent space, become simplices of the space.
For example, take the MALL formula $\mathrm{Bool}:=1\oplus 1$ and write $\mathrm{Bool}^{\& k}:=\mathrm{Bool}\,\&\cdots\&\,\mathrm{Bool}$ ($k$ times).
Let us consider first $\mathrm{Bool}^{\& 2}$. If we call $\set{T_1,F_1,T_2,F_2}$ its web, then $\llbracket\mathrm{Bool}^{\& 2}\rrbracket$ is the graph (in the usual sense that the web is given by the vertices and the simplices by the cliques) in Figure~\ref{fig:Bool2}.
When realized geometrically, we get the circle $S^1$ (modulo homeomorphism).
Now take $\mathrm{Bool}^{\& 3}$. If we call $\set{T_1,F_1,T_2,F_2,T_3,F_3}$ its web, then $\llbracket\mathrm{Bool}^{\& 3}\rrbracket$ is given by the graph in Figure~\ref{fig:Bool3}.
The maximal cliques of this graph are the  $8$ ``faces of the octahedron'', so when realized geometrically we get the sphere $S^2$ (modulo homeomorphism).
Observe that in the previous examples, \emph{all} the simplices of the asc $\llbracket\mathrm{Bool}^{\& k}\rrbracket$ are ``witnessed'' by a proof of $\vdash \mathrm{Bool}^{\& k}$, in the sense that any simplex is contained in some proof. This is due to the fact that the maximal cliques are exactly the interpretation of some proof.
For instance, for $k=3$, it is easy to see that in MALL there are exactly the $2^3=8$ cut-free proofs of $\mathrm{Bool}^{\& 3}$ given in Figure~\ref{fig:piijm}, for $i,j,m=1,2$.
Now it is easily seen that, setting $a^h_1=T_h$ and $a^h_2=F_h$ for $h=1,2,3$, one has $\llbracket \pi_{ijm} \rrbracket=\set{a^1_i,a^2_j,a^3_m}$, that is, the $8$ faces of the octahedron.\\
However, the fact that all the cliques are witnesses by a proof (in the previous sense) is due to the simple cases we have considered. In fact, it is not always the case:
take for instance ``Gustave's formula'' $G:=(1 \oplus (1 \& 1)) \parr (1 \oplus (1 \& 1)) \parr (1 \oplus (1 \& 1))$ of MALL.
If we call $\set{\bot,T,F}$ the web of $\llbracket 1 \oplus (1 \& 1) \rrbracket$, then $\llbracket G \rrbracket$ has web $\web G=\set{\bot,T,F}^3$.
Let us call $a_1:=(\bot,T,F),a_2:=(F,\bot,T),a_3:=(T,F,\bot)\in\web G$.
Now, it is easily seen that there are MALL proofs $\pi_{1,2,3}: \, \vdash 1 \oplus (1 \& 1)$ s.t.\ $\llbracket \pi_1 \rrbracket\supseteq\set{a_2,a_3}$, $\llbracket \pi_2 \rrbracket\supseteq\set{a_1,a_3}$ and $\llbracket \pi_3 \rrbracket\supseteq\set{a_1,a_2}$.
So each of the previous three sets is an edge of $\llbracket G \rrbracket$, and the definition of coherence space forces thus $x:=\set{a_1,a_2,a_3}$ to be a $2$-simplex of $\llbracket G \rrbracket$.
But it can be proven that there is \emph{no} MALL proof of $G$ whose coherent semantics contains the whole $x$.\\
Differently said, the geometry of the asc $\llbracket A \rrbracket$ in general does not reflect the existence of proofs of $A$, since coherent spaces ``fill'' unnecessary $n$-holes: there are simplices that do not come from proofs.
In order to have a faithful \emph{geometrical representation of the (interpretation of the) proofs}, we propose to consider the sub-asc $[A]$ of $\llbracket A \rrbracket$ given by the web $\web{[A]}:=\web A$ and simplices $\mathcal{S}[A]:=\set{x\subseteq \llbracket \pi \rrbracket \, \st \pi :\, \vdash A}$.
In $[A]$, every simplex does come from a proof, in the sense that it is a face of the interpretation of a proof (i.e.\ it is contained in it).
Remark that the construction of $[A]$ makes sense if $\llbracket \cdot \rrbracket$ is any ``webbed semantics'' of LL, not just the coherent one. We think of $[A]$ as the ``canonical'' \emph{geometrical} object associated with any such semantics $\llbracket \cdot \rrbracket$.
Remark also that $[A]$ is, of course, determined by the formula $A$, but only in the fact that (for a fixed system -- here we took MALL) a formula determines its proofs.
More precisely, $[A]$ provides the geometrical organisation of the interpretations of the proofs of $A$ as an asc (so a pedantic notation for $[A]$ would be something like $[\mathrm{Proofs}_{\llbracket\cdot\rrbracket}(A)]$).
Our aim is to relate, if possible, geometrical properties of the asc $[A]$ with proof-theoretical/computational properties of $A$ (read: of the proofs of $A$).
For instance, the geometrical property of having a ``$n$-hole'' (the presence of a non-simplex), is determined by the absence of suited proofs (as it happens for $[G]$, in contrast to what we have seen for $\llbracket G\rrbracket$).
Until now our ``running example'' was the coherent semantics; from now on it will be the relational semantics.

A tool to study the $n$-holes of an asc is provided by (simplicial) homology, which we quickly recall:

\begin{definition}
 
 A \emph{chain complex}\index{Chain!-complex} on $\mathbb{Z}$ is the data of a sequence of $\mathbb{Z}$-modules $M_k$ together with linear maps
 $
  \cdots\xrightarrow{\partial_{k+2}}
  M_{k+1}\xrightarrow{\partial_{k+1}}
  M_k \xrightarrow{\partial_k}
  M_{k-1} \xrightarrow{\partial_{k-1}}\cdots
 $
 such that
$\partial_{k}\circ \partial_{k+1}=0$.
 
\end{definition}

\begin{definition}
 Let $(M_k,\partial_k)_{k\in\N}$ be a chain complex.
 Elements of $\mathrm{Im}(\partial_{k+1})$ are called \emph{$k$-boundaries} and elements of $\mathrm{Ker}(\partial_k)$ are called \emph{$k$-cycles}. Both are sub-$\mathbb{Z}$-modules of $M_k$. By definition of chain complex, a $k$-boundary is a $k$-cycle, so the (abelian) group structure of $\mathrm{Im}(\partial_{k+1})$ forms a subgroup (thus, normal) of $\mathrm{Ker}(\partial_k)$, and we can take the abelian quotient group
 $
  \mathcal{H}_k:= \mathrm{Ker}(\partial_k)/\mathrm{Im}(\partial_{k+1})
 $
 which is called the \emph{$k$-homology group} of $(M_k,\partial_k)_{k\in\N}$.
 Its elements are called \emph{$k$-homology calsses on $\mathbb{Z}$} and are the cosets $\gamma + \mathrm{Im}(\partial_{k+1})$ for $\gamma \in \mathrm{Ker}(\partial_k)$.
 The group $\mathcal{H}_k$ inherits a structure of $\mathbb{Z}$-module via the multiplication in $\mathrm{Ker}(\partial_k)$.
\end{definition}

\begin{definition}
 
 A \emph{chain map} from a chain complex $(M_k,\partial_k)_{k\in\N}$ to a chain complex $(M'_k,\partial'_k)_{k\in\N}$ is the data of a sequence of linear maps $\varphi_k:M_{k}\rightarrow M'_k$ commuting with borders, that is, such that one has the commutative diagrams:
$\varphi_{k}\circ \partial_{k+1}=\partial'_{k+1}\circ \varphi_{k+1}$
 Chain complexes on $\mathbb{Z}$ with chain maps as morphisms form a category, denoted $\mathrm{Chain}_\mathbb{Z}$.
 
\end{definition}

\begin{remark}

 Every chain map $\varphi=(\varphi_k)_{k\in\N}$ from a chain complex $M=(M_k,\partial_k)_{k\in\N}$ to a chain complex $M'=(M'_k,\partial'_k)_{k\in\N}$ induces linear maps $\mathcal{H}_k \varphi:\mathcal{H}_k(M)\rightarrow \mathcal{H}_k(M')$ setting, for all $k$-cycles $\gamma$ in $M$,
  $
  \mathcal{H}_k \varphi (\gamma + \mathrm{Im}(\partial_{k+1})):= \varphi_{k}(\gamma) + \mathrm{Im}(\partial'_{k+1}).
  $
  We have thus a functor
  $
  \mathcal{H}_k:\mathrm{Chain}_\mathbb{Z}\rightarrow \mathrm{Modules}_\mathbb{Z}
  $
(where the latter is the category is of $\mathbb{Z}$-modules and linear maps).
 
\end{remark}

\begin{remark}\label{lm:HisFunctor}

Any functor $F$ from a category $\mathcal{A}$ to $\mathrm{Chain}_\mathbb{Z}$ induces functors 
$
  \mathcal{H}_k\circ F:\mathcal{A}\rightarrow~\mathrm{Modules}_\mathbb{Z}.
 $
In algebraic topology one defines the \emph{chain-modules} $\mathcal{C}_k X$ of the \emph{oriented $k$-simplices} and \emph{boundary operators} $\partial_{k+1}: \mathcal{C}_{k+1} X \rightarrow \mathcal{C}_k X$ associated with an asc $X$.
In the following, we consider these constructions known by the reader (otherwise, see \cite{Hatcher}).
Those form a chain-complex $\mathcal{C}X:=(\mathcal{C}_k X,\partial_{k})_{k\in\N}$ on $\mathbb{Z}$, which extends to a functor $\mathcal{C}:\mathrm{ASC}\rightarrow \mathrm{Chain}_\mathbb{Z}$.
We can thus consider functors $\mathcal{H}_k\circ \mathcal{C} :\mathrm{ASC}\rightarrow \mathrm{Modules}_\mathbb{Z}$ (take $\mathcal A:=\mathrm{ASC}$ and $F:=\mathcal C$) obtaining the \emph{$k$-homology $\mathbb{Z}$-modules} $\homology{k}{X}$ of an asc $X$.
\end{remark}

In the framework of denotational semantics, however, one is typically led to consider a ``relational based'' notion of morphisms, rather than a functional one -- for instance, the relational semantics interprets a proof as a (simplicial) \emph{relation}, not as a simplicial map.
The question is, thus, what kind of morphism should we consider between the asc's ?
Said with the notation of Remark \ref{lm:HisFunctor}, what should we take as our category $\mathcal A$ and functor $F$?
A natural request the answer should satisfy, would be to make the homology modules be type-isomorphism invariants (two formulas $A,B$ are \emph{type-isomorphic} iff there are proofs $\pi:\, A\vdash B$ and $\pi^{-1}:\, B\vdash A$ s.t.\ $\textit{cut}(\pi,\pi^{-1})\cutelim\textit{ax}_A$ and $\textit{cut}(\pi^{-1},\pi)\cutelim\textit{ax}_B$).
This happens to be a non-trivial problem: in the following we show how some ``direct'' tries fail, and how we can finally achieve this property, but with a drawback.

\section{Hints on the possible solutions}\label{sec:hints}

The first idea is to take $\mathcal A:=\mathrm{RelASC}$, the category whose objects are the asc's and whose morphisms are the \emph{simplicial relations}, that is, the elements of $\mathrm{RelASC}(X,Y)$ are the $t\subseteq\web{X}\times\web{Y}$ s.t.\ $t\cdot\sigma \in\simplices{Y}$ for all $\sigma\in\simplices{X}$, where for a $\sigma\subseteq\Pfinstar{\web{X}}$ we set: $t\cdot\sigma:=\bigcup_{a\in\sigma} t^+(a)$ and $t^+:\web{X}\rightarrow \mathscr{P}(\web{Y})$ is given by $t^+(a):=\set{b\in\web{Y}\,\,\st (a,b)\in t}$.
This is because the relational semantics interprets a type $A$ as a set $\llbracket A \rrbracket$ and a proof $\pi: \, \vdash A\multimap B$ as a subset $\llbracket \pi \rrbracket\subseteq \llbracket A \rrbracket\times \llbracket B \rrbracket$ satisfying the following ``simplicial property'':
 $\llbracket \pi \rrbracket\cdot\sigma \in [B]$ for all $\sigma\in [A]$.
The natural choice for $F$ now would be $FX:=\mathcal{C}X$, $Ft:=(F_k t)_k$ with $F_k t: \mathcal{C}_k X\rightarrow \mathcal{C}_k Y$ given by extending by linearity the association:
$
 F_k t(a_0,\dots,a_k):=\sum
 (b_0,\dots,b_k)
$
where 
the sum is on all the \emph{oriented simplices} (see \cite{Hatcher}) $(b_0,\dots,b_k)$ s.t.\ $(a_i,b_i)\in t$ for $i=0,\dots,k$.
However, such $Ft$ is not a chain map (it does not commute with borders), and $F$ is not even functorial from $\mathrm{RelASC}$ to $\mathrm{Chain}_\mathbb{Z}$.
Already one would suffice to invalidate the construction, but let us show both.\\
- \emph{Non} chain map: take $X=S^1$ (also $X=\Delta^2$ works) with web $\set{a_0,a_1,a_2}$, $Y=\Delta^2$ with web $\set{b_0,b_1,b_2}$, and $t\in\mathrm{RelASC}(X,Y)$ given by the set $\set{(a_0,b_0),(a_0,b_1),(a_1,b_1),(a_2,b_2)}$.
Now it is easily seen that, for instance, $\partial_1(F_1 t(a_0,a_2))=2(b_2)-(b_0)-(b_1)
\neq
(b_2)-(b_0)-(b_1) = F_0 t(\partial_1(a_0,a_2))$.\\
- \emph{Non} functoriality: take asc's $X$,$Y$,$Z$ with $X$ containing a vertex $a$, $Y$ an edge $\set{b,b'}$ and $Z$ a vertex $c$. Now take $s:=\set{(a,b),(a,b')}\in\mathrm{RelASC}(X,Y)$ and $t:=\set{(b,c),(b',c)}\in\mathrm{RelASC}(Y,Z)$.
We have: $F_0(t\circ s)(a)=(c)\neq2(c)=F_0 t(F_0 s(a))$.

The problems seem to be related to the fact the relational semantics is qualitative and not quantitative. So we could think to consider a quantitative one, such as the matrix semantics. It interprets a type $A$ as the relational semantics, but a proof $\pi:\, \vdash A\multimap B$ as a $\mathbb{N}$-matrix on $\mid\!\!A\!\!\mid\times \mid\!\!B\!\!\mid$, that is, a function $t:\, \mid\!\!A\!\!\mid\times \mid\!\!B\!\!\mid\rightarrow \mathbb{N}$. Let us write $t_{ab}$ for $t(a,b)$ and $t_a$ for the multiset $t_a:\,\mid\!\!B\!\!\mid\rightarrow\mathbb{N}$, $t_a(b):=t_{ab}$.
Now we could think to consider the category $\mathcal A:=\mathrm{MatrixASC}$ of asc's and morphisms from $X$ to $Y$ the matrices $t$ on $\web X\times \web Y$ s.t.\ $\bigcup\limits_{a\in\sigma} \textit{support}(t_a)\in \simplices Y$ for all $\sigma\in \simplices X$.
The natural choice for $F$ now would be as before, but by taking into account coefficients in the action on the morphisms, setting:
$
 F_k t(a_0,\dots,a_k):=\sum
 \left(\prod\limits_{i=0}^k t_{a_i b_i}\right)
 (b_0,\dots,b_k).
$
However also this fails to form a chain map. In fact, take $X=Y=S^1$ (or also $\Delta^1$ would work) with vertices $\set{a_0,a_1,a_2}$, and take $t\in\mathrm{MatrixASC}(X,X)$ the matrix in Figure~\ref{fig:matrix}.
It is a simple verification that one has:
$\partial_1(F_1 t(a_0,a_1))=\partial_1(a_1,a_2)=(a_2)-(a_1)
\neq
-(a_0)=(t_{a_1 a_0}-t_{a_0 a_0})(a_0)+(t_{a_1 a_1}-t_{a_0 a_1})(a_1)+(t_{a_1 a_2}-t_{a_0 a_2})(a_2)=F_0 t(\partial_1(a_0,a_1))$.

Since also the previous construction did not work, we could now try to impose some constraint on the morphisms, by considering the more exotic setting of \emph{coherence asc's}, that is, the data $X$ of an asc $\underline{X}=(\web{X},\mathcal{S}X)$ and of a \emph{cs} $\mathrm{cs}(X)=(\web{X},\smallfrown_X)$ (on the same web).
 It is easily checked that coherence asc's form a category
, where the morphisms from $X$ to $Y$ are the cliques $t\subseteq\web{X}\times\web{Y}$ of the \emph{cs} $\cs{X}\multimap\cs{Y}$ that are also simplicial relations from $\underline X$ to $\underline Y$.
In this setting, the natural way to define $F$ is as in the case of relational semantics. But this fails again to be a chain map, because we can reproduce the same exact counterexample: take $X$ the asc $\underline X =S^1$ with the trivial coherence -- only the singletons are coherent with themselves -- and take $Y$ the asc $\underline{Y} = \Delta^2$ with the trivial coherence plus $b_0\frown b_1$. Now the same exact $t$ of the previous case is a counterexample for the commutation with the borders.

\begin{figure}[t]
    \centering
    \begin{minipage}{0.5\textwidth}
        \centering
         \begin{footnotesize}\begin{tabular}{c|c|c|c|}
      & $a_0$ & $a_1$ & $a_2$ \\
\hline
  $a_0$ &  $1$  &  $0$  &  $1$  \\
\hline
  $a_1$ &  $0$  &  $0$  &  $1$  \\
\hline
  $a_2$ &  $1$  &  $0$  &  $0$  \\
\hline
\end{tabular}\end{footnotesize}
        \caption{The counterexample matrix}\label{fig:matrix}
    \end{minipage}
    \begin{minipage}{0.5\textwidth}
        \centering
        \begin{footnotesize}\begin {tikzpicture}
\fill[gray!70] (13.25,1) -- (12.35,0.2) -- (12.25,-1) -- cycle;
\fill[gray!70] (13.25,1) -- (14.15,0.2) -- (14.25,-1) -- cycle;
\fill[gray!70] (12.25,-1) -- (13.25,-1.36) -- (14.25,-1) -- cycle;
\filldraw[black] (12.25,-1) circle (1pt)
node[anchor=north] {$\set{a}$};
\filldraw[black] (14.25,-1) circle (1pt) node[anchor=north] {$\set{b}$};
\filldraw[black] (13.25,1) circle (1pt) node[anchor=west] {$\set{c}$};
\path[red] (12.25,-1) edge (14.25,-1);
\path[red] (13.25,1) edge (14.25,-1);
\path[red] (12.25,-1) edge (13.25,1);
\filldraw[black] (12.35,0.2) circle (1pt)
node[anchor=south east] {$\set{a,c}$};
\filldraw[black] (14.15,0.2) circle (1pt)
node[anchor=south west] {$\set{b,c}$};
\filldraw[black] (13.25,-1.36) circle (1pt)
node[anchor=north] {$\set{a,b}$};
\path (13.25,1) edge (12.35,0.2);
\path (12.35,0.2) edge (12.25,-1);
\path (13.25,1) edge (14.15,0.2);
\path (14.15,0.2) edge (14.25,-1);
\path (12.25,-1) edge (13.25,-1.36);
\path (13.25,-1.36) edge (14.25,-1);
        \end{tikzpicture}\end{footnotesize}
        \caption{$S^1$ in red and $\barI{S^1}$ in black}\label{fig:IS1}
    \end{minipage}
\end{figure}


We show now our (partial) solution. We consider again the category $\mathcal A := \mathrm{RelASC}$, we transform a simplicial relation into a simplicial map, and finally we rely on the standard topological construction of $\mathcal C$ on $\mathrm{ASC}$ in order to get a homology functor.
The price to pay is that, in such transformation, we also transform the considered asc.

\begin{definition}

 We define the following endofunctor $\mathscr{I}$ on $\mathrm{ASC}$.
 If $X$ is an asc, then $\mathscr{I}X$ is the asc with web $\web{\mathscr{I}X}\,:=\simplices{X}$ and $k$-simplices the sets $\set{u_0,\dots,u_k}\subseteq^*_{\mathrm{fin}}\simplices{X}$ s.t.\ $\bigcup_{i=0}^k u_i\in\simplices{X}$.
 If $f:\web X \rightarrow \web Y$ is a simplicial map from $X$ to $Y$, then $\mathscr{I}\!f:\simplices{X}\rightarrow \simplices{Y}$ is given by the direct image: $\mathscr{I0\!}f(u):=fu$.

\end{definition}

\begin{proposition}
 
 The endofunctor $\mathscr{I}$ forms a monad on $\mathrm{ASC}$ with unit $\varepsilon:\mathrm{id}\Rightarrow \mathscr{I}$ and multiplication $\mu: \mathscr{I}^2\Rightarrow \mathscr{I}$ whose components are respectively:
 $\varepsilon_X:\web X \rightarrow \simplices{X}$ given by $\varepsilon_X(a):=\set{a}$, and $\mu_X: \simplices{\mathscr{I}\!X} \rightarrow \simplices{X}$ given by $\mu_X(U):=\bigcup\limits_{u\in U} u$.
\end{proposition}

\begin{proposition}
 
 There is an equivalece of categories between $\mathrm{RelASC}$ and the Kleisli category $\mathrm{ASC}_{\!\mathscr{I}}$ of $\mathscr I$, which is given by the functor
 $
  (\cdot)^+: \mathrm{RelASC}\rightarrow \mathrm{ASC}_{\!\mathscr{I}}
 $
 defined by $X^+:=X$ and, for a simplicial relation $t\subseteq \web X \times \web Y$, $t^+\subseteq \web X \times \simplices{Y}$ is the function $t^+$ already defined in the first lines of Section \ref{sec:hints}.

\end{proposition}

We have thus the following functors and, with the notation of Remark \ref{lm:HisFunctor}, we are taking $F=\mathcal C \circ R_{\!\!\mathscr{I}} \circ (\cdot)^+$. 
 \[
 \begin{tikzcd}
  \mathrm{RelASC} \arrow[r, "(\cdot)^+", "\cong"']
  & \mathrm{ASC}_{\!\!\mathscr{I}} \arrow[r, "R_{\!\!\mathscr{I}}"]
  & \mathrm{ASC} \arrow[r, "\mathcal{C}"]
  & \mathrm{Chain}_{\mathbb{Z}} \arrow[r, "\mathcal{H}_k"]
  & \mathrm{Modules}_\mathbb{Z}
 \end{tikzcd}
 \]
where $R_{\!\!\mathscr{I}}$ is part of the Kleisli adjunction $L_{\!\!\mathscr{I}} \dashv R_{\!\!\mathscr{I}}$. That is, $R_{\!\!\mathscr{I}}X:=\mathscr{I}\!X$ and $R_{\!\!\mathscr{I}}f:=\mu_Y\circ \mathscr{I}\!f$.
 
Now, by using the relational semantics $\llbracket .\rrbracket$ interpretation functor, we get:

\begin{corollary}
 
 If $A$ and $B$ are isomorphic types, then $\mathcal{H}_k(\barI{\,[\!A]})\cong\mathcal{H}_k(\barI{\,[\!B]})$.
 
\end{corollary}
 
\paragraph{Comments and questions}

We declared that our interest is the study of the geometry of the asc $[A]$, and we tackled it by regarding its homology. 
Because of the constraint to take simplicial relations as morphisms instead of simplicial maps, we find ourself able to talk in a satisfactory (functorial) way only about the modified asc $\barI{\,[\!A]}$.
Therefore the question is: what is the relation between $\barI{X}$ and $X$, or at least between the geometry of $\barI{X}$ and that of $X$, for whatever ``geometry'' could mean ?
It is immediate to check that $\barI{\Delta^1}=\Delta^2$. 
Therefore $\barI{X}$ makes the dimension grow and thus, in general, the geometric realizations are not homeomorphic.
Figure~\ref{fig:IS1} shows another interesting example.
We remark that, even if their geometric realizations are not homeomorphic, for the above examples we still have that $\mathrm{Geo}(\barI X)$ \emph{retracts onto} $\mathrm{Geo}(X)$. Therefore, in this very simple cases, the homologies of $X$ and of $\barI X$ remain the same.
Is this always the case?
If not, we think that the case $\barI{S^2}$ would be enough to find a counterexample (but drawing it by hand seems already too complicated; maybe softwares such as \emph{SAGE} could help here).
Is the transformation $\barI{}$ meaningful from a geometric point of view?
Is it from a logical/computational one? 
\bibliographystyle{eptcs}
\bibliography{MyBibTeX}

\end{document}